\documentclass[
twocolumn,prl,showpacs,preprintnumbers,amsmath,amssymb]{revtex4}

\usepackage{graphicx}
\usepackage{amssymb}
\usepackage{amsmath}
\usepackage{dsfont}
\usepackage{bm}
\usepackage{mathrsfs}

\usepackage{graphicx}

\begin{document}

\title{Electromagnetically induced invisibility cloaking}
\author{Darran F. Milne and Natalia Korolkova}
\affiliation{
School of Physics and Astronomy, University of St Andrews,
North Haugh, St Andrews KY16 9SS, UK}
\date{\today}
\begin{abstract}
Invisibility cloaking imposes strict conditions on the refractive index profiles of cloaking media that must be satisfied to successfully hide an object. The first experimental demonstrations of cloaking used artificial metamaterials to respond to this challenge. In this work we show how a much simpler technique of electromagnetically induced transparency can be used to achieve a partial, {\it carpet} cloaking at optical frequencies in atomic vapours or solids. To generate a desired combination of low absorption with strong modifications of the refractive index, we use chiral media with an induced magneto-electrical cross-coupling. We demonstrate that high-contrast positive refractive indices can be attained by fine tuning the material with a gradient magnetic field and calculate the parameters required to construct a carpet cloak.
\end{abstract}
\maketitle

In electromagnetically induced transparency (EIT) \cite{EIT} opaque substances are made transparent by external electromagnetic fields. The fields prepare the substance in dark states \cite{EIT} that are decoupled from light of a certain frequency and polarization, thus making the material transparent for such light. EIT has been the basis for slow light in atomic vapors \cite{SlowAtom,slowatom2} or solids \cite{SlowSolid}. Here we put forward a method of electromagnetically induced cloaking where external fields manipulate the refractive index of a specially designed EIT material to guide light around an object without making the object itself transparent. Electromagnetically induced transparency and slow light have inspired applications in metamaterials \cite{MetaEIT,Rainbow}; here we adopt a concept that originated in the area of metamaterials --- cloaking \cite{Greenleaf,LeoConform,PSS,CCS,Book} --- and show how it can be implemented in atomic vapors and solids.

A cloaking device consists of a transparent mantel that encloses the object to be hidden. The mantel refracts light around its hidden interior such that light exits as if it had traversed empty or uniform space, thus hiding the object and the act of hiding itself. For this, certain refractive-index profiles are required that are difficult to implement in ordinary optical materials but are (barely) feasible in metamaterials \cite{Schurig}. In most cases, these should be optically anisotropic materials, but it is also possible to cloak with isotropic materials \cite{LeoConform}. Electromagnetically induced transparency is able to create index profiles suitable for lensing \cite{Dunn} but with an index contrast that is too low for cloaking. Here we point out that another method \cite{Fleisch,Sikes} borrowed from metamaterials research --- the implementation of chiral materials \cite{Chiral} --- is capable of cloaking. The currently achievable index contrast is only sufficient for partial cloaking, also known as ``carpet cloaking'' \cite{LiPendry}, where a curved object is made to appear flat. The flattened object can then be camouflaged against the background to render it effectively invisible.

The index profiles for cloaking are often designed by the optical implementation of coordinate transformations \cite{Book}. Here, the cloaking device creates the illusion that physical space is transformed into a virtual space, which is most easily explained by a two-dimensional example \cite{LeoConform} that also serves to introduce carpet cloaking. Suppose that light propagates in a planar material. We describe the Cartesian coordinates $x$ and $y$ of the physical propagation plane with the complex number $z=x+iy$. Imagine we transform $z$ to the new complex coordinate $z'$ of a virtual plane where $z'$ is only a function of $z$ and not of $z^*$ (such a function $z'(z)$ is analytic and defines a conformal map \cite{Needham}). Suppose that the virtual plane is empty. In this case light experiences a virtual geometry described by the line element $dl'$. In the physical plane, $dl'$ appears as $|dz'/dz|\,dl$. Now, according to Fermat's principle \cite{Book}, in an optically isotropic medium with refractive-index profile $n$ light rays follow geodesics with the path length measured by $n\,dl$. Therefore, if we identify $n$ with $|dz'/dz|\,dl$, the medium implements the coordinate transformation $z'(z)$. Consider, for example the Zhukowski transform
\begin{equation}
z' = z+\frac{a^2}{z} \,,\quad z = \frac{z'\pm\sqrt{z'^2-4a^2}}{2}
\label{zhu}
\end{equation}
with constant $a$ (defining the characteristic scale of the device) that is implemented by the refractive-index profile
\begin{equation} 
n = \left|\frac{dz'}{dz}\right| =  \left|1-\frac{a^2}{z^2}\right|\,.
\label{n}
\end{equation}
\begin{figure}
\includegraphics[width=8cm]{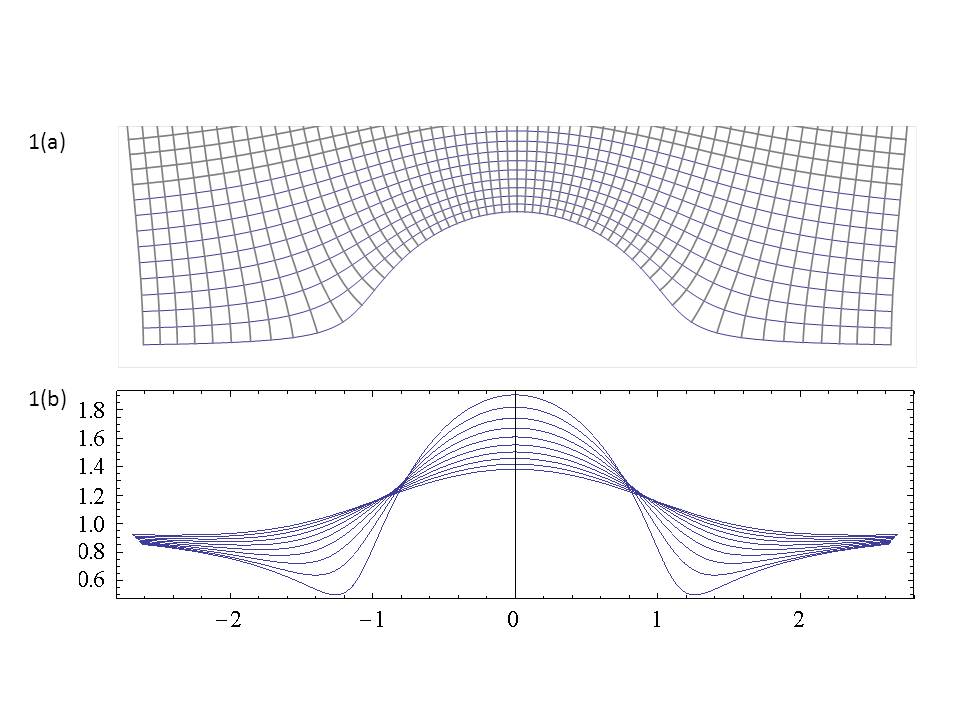}
\caption{(a) The grid shows the curved path of light in the physical space $z$ due to the transformation in Eq.(\ref{zhu}). (b) The variation of refractive index at the boundary for some values of $\eta$. }
\label{map}
\end{figure}
Fig.~\ref{map}(a) shows that straight lines in virtual space are curved lines in physical space, in particular the lines $z'=x+i\eta$ with $\eta=\mathrm{const}$. As light experiences the geometry of the virtual space, such curved lines appear to be flat and so does any object behind the curve $z(x+i\eta)$. An optical material with the index profile of Eq.~(\ref{n}) above the curve $z=(x+i\eta)$ thus optically flattens this curve and anything underneath, which is the defining property of carpet cloaking \cite{LiPendry}. Instead of conformal maps one could use quasi-conformal transformations \cite{LiPendry} that have the advantage of acting in a finite domain. The strongest deformations of the conformal map occur at the boundaries and thus require the highest refractive-index contrast there. 
Figure \ref{map}(b) shows the index profiles at the boundaries  $z=(x+i\eta)$ for various values of $\eta$. In the following we discuss how such optical properties can be electromagnetically induced.

We want to take advantage of the quantum coherence properties of atomic media exploiting electromagnetically induced transparency to manage the refractive index profile and, at the same time, to profit from the low absorption window. Inspired by the lensing effects demonstrated in standard EIT \cite{Dunn}, we seek to enhance the (positive) refractive index contrast by introducing  magneto-electric cross-coupling. Recently, \cite{Chiral,Fleisch} have shown that such chiral media can be manipulated to attain negative refractive indices. These schemes have been further improved upon in \cite{Sikes} using off resonant Raman transitions. To avoid the inherent experimental challenges associated with decoherence in atomic vapors we suggest a solid state implementation of chiral EIT media \cite{SlowSolid}.
 
\begin{figure}
\includegraphics[width=7.5cm]{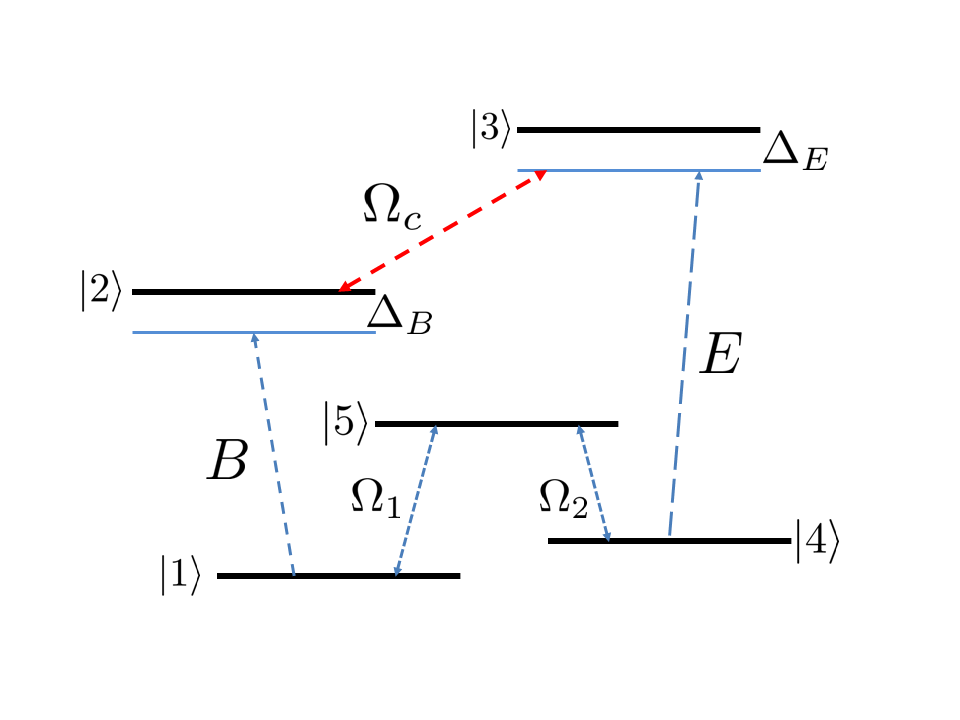}
\caption{A 5-level scheme for the implementation of tunable refraction via electromagnetically induced cross coupling. The magnetic dipole transition $|2\rangle$ - $|1\rangle$ and the electric dipole transition $|3\rangle$-$|4\rangle$ are coupled by $\Omega_c$ to induce chirality. The "ground" state of the system is formed by the dark state $|D\rangle$ of the subsystem $\{|1\rangle, |4\rangle, |5\rangle\}$. For details see \cite{Fleisch}. }
\label{5level}
\end{figure}

{\it EIT in chiral media}.
The electromagnetic constitutive equation relations between medium polarization \textbf{P} or magnetization \textbf{M} and electromagnetic fields \textbf{E} and \textbf{H} are usually expressed in terms of permittivity and permeability only. They can however be generalized to include the \textit{chirality} effects, i.e., the cross coupling between the electric and magnetic fields:
\begin{eqnarray}\label{pol}
\textbf{P}= \overline{\chi}_{e} \textbf{E} + \frac{\overline{\xi}_{EH}}{4\pi} \textbf{H},\\
\label{mag}
\textbf{M}=  \frac{\overline{\xi}_{HE}}{4\pi} \textbf{E} + \overline{\chi}_{m} \textbf{H},
\end{eqnarray}
where $\overline{\epsilon} = 1+4\pi \overline{\chi}_{e}$ and $\overline{\mu} = 1+4\pi \overline{\chi}_{m}$ are the complex valued permittivity and permeability tensors. These generalized constitutive equations describe media with \textit{magneto-electric cross-coupling} which are also known as bianisotropic media. $\overline{\xi}_{EH}$ and $\overline{\xi}_{HE}$ denote tensorial coupling coefficients between the electric and magnetic degrees of freedom. We can imagine a simple chiral medium to be composed of three level atomic systems, with ground states $|g\rangle$ and two excited states $|1\rangle$ and $|2\rangle$. If we choose $|g\rangle \rightarrow |1\rangle$ to be a magnetic dipole transition and $|g\rangle \rightarrow |2\rangle$ to be an electric dipole transition, we can induce cross coupling through a strong resonant coupling of the upper two states. Then the magnetization given by the transition from the ground state to $|1\rangle$ will contain an additional contribution from the electric field and vice versa for the polarization. Note, that such a simple 3-level system is resemblant of a typical $\lambda$-scheme for EIT. Such a medium can be realized by inducing chirality in a solid state EIT system, using doped bulk crystals \cite{SlowSolid}. In practice, an extended 5-level atomic system should be used (Fig.~\ref{5level}), which will be discussed later.

To find the refractive index, we examine the propagation properties of the electromagnetic waves in such media as governed by the Helmholtz equation:
\begin{equation}
[\overline{\epsilon} +(\overline{\xi}_{EH} + \frac{c}{\omega} \textbf{k} \times )\overline{\mu}^{-1} (\frac{c}{\omega}\textbf{k} \times -\overline{\xi}_{HE})]\textbf{E}=0.
\end{equation}
The solution to this equation for the wavevector \textbf{k} is in general very complicated \cite{helmholz}. The analysis can be simplified by assuming the permittivity and the permeability to be isotropic, $\overline{\epsilon}=\epsilon \mathbb{1}$, $\overline{\mu}=\mu \mathbb{1}$. Further, we confine our analysis to the 1D theory with the wave propagating in the $z$-direction. Finally, we only consider media which allow for conservation of the photonic angular momentum at their interfaces. This leads to the following expression for the refractive index (see also \cite{Fleisch}):
\begin{equation}\label{index}
n^{+}=\sqrt{\epsilon \mu - \frac{(\xi^{+}_{EH} + \xi^{+}_{HE})^{2}}{4}} + \frac{i}{2} (\xi^{+}_{EH} - \xi^{+}_{HE}).
\end{equation}
Note that Eq.(\ref{index}) has more degrees of freedom than the non-chiral version and allows for a tunable refractive index that can even by allowed to go negative without requiring a negative permeability. For $\xi >0$, Eq. (\ref{index}) can be further simplified by setting $\xi^{+}_{EH} = - \xi^{+}_{HE} = i \xi$ thus giving $n=\sqrt{\epsilon \mu} - \xi$. Then adaptive changes to the refractive index in a chiral media are performed by altering the magnitudes of the coupling coefficients $\xi_{EH}$ and $\xi_{HE}$. These are dependent on the strength of the coupling field $\Omega_C$ and the detuning from the magnetic and electric energy levels $\Delta$ \cite{Fleisch,Chiral}.

\textit{Managing the refractive index of a chiral medium}.
Using the theory of the chiral electromagnetic media developed in \cite{Fleisch}, we have identified the parameters of the system to facilitate the refractive index profile desired for invisibility cloaking.  Our goal is to tailor the refractive index of such a material in acordance with Fig.~\ref{map} (b), controlling the electric and magnetic dipole levels $\Delta_E$ and $\Delta_B$ and the cross coupling Rabi frequency $\Omega_c$.

Consider the system modeled by the 5-level atomic system interacting with the strong laser beams $\Omega_{1,2}$ and the cross-coupling light field $\Omega_c$ as presented in Figure~\ref{5level}. The transition $|2\rangle - |1\rangle$ is a magnetic dipole transition in nature. All others are electric dipole transitions. The Hamiltonian $H=H_{atom} - d.E(t)-\mu .B(t)$ of the system can be expressed in a convenient matrix form:
\begin{widetext}
\begin{eqnarray}
H= \left( \begin{array}{ccccc}\label{hmat}
\hbar \omega_1 & -\frac{1}{2} \mu_{21} B e^{i\omega_p t} & 0 & 0 & -\frac{\hbar}{2}\Omega_1 e^{i \omega_1 t} \\
-\frac{1}{2} \mu_{21} B e^{-i\omega_p t} & \hbar \omega_2 & -\frac{\hbar}{2}\Omega^{*}_c e^{i \omega_c t} & 0 & 0 \\
0 & -\frac{\hbar}{2}\Omega_c e^{-i \omega_c t} & \hbar \omega_3 & -\frac{1}{2}d_{34} E e^{-i \omega_p t} & 0 \\
0 & 0 & -\frac{1}{2}d_{34} E e^{i \omega_p t} & \hbar \omega_4 & -\frac{\hbar}{2} \Omega_2 e^{i\omega_2 t} \\
-\frac{\hbar}{2}\Omega_1 e^{-i \omega_1 t} & 0 & 0 & \frac{\hbar}{2} \Omega_2 e^{-i \omega_2 t} & \hbar \omega_5
\end{array} \right)\end{eqnarray}
\end{widetext}
with the electric and magnetic dipole moments 
\begin{equation}
d_{34} = \langle 3|er.\hat{e}_{E} |4\rangle ,\;\;\; \mu_{21} = \langle 2|\mu.\hat{e}_{B} |1\rangle.
\end{equation}
$E$ and $B$ are the electric and magnetic components of the weak probe field which oscillates at a frequency $\omega_{p}$. The Rabi frequencies $\Omega_{1}$, $\Omega_{2}$ and $\Omega_{c}$ belong to strong coupling lasers which oscillate at frequencies $\omega_{1}$, $\omega_{2}$ and $\omega_{c}$ respectively. We choose $d_{34}$, $\mu_{21}$, $\Omega_{1}$ and $\Omega_{2}$ to be real but allow the strong field $\Omega_{c}$ to stay complex. Using the Liouville equation, we have found the equations of motion for all twenty five density matrix elements. This system of equations is simplified considerably by focusing on just the linear response so we treat probe fields $E$ and $B$ as weak fields which allows us to neglect the upper state populations $\rho_{33}$ and $\rho_{22}$. 

The true index of refraction which takes the full tensor form i.e. the angle dependent correction to $\epsilon$ and $\mu$, gets very complicated. However under the assumption of isotropic permittivity and permeability we can neglect the tensor form for the electric and magnetic polarizabilities $\alpha^{EE}$ and $\alpha^{BB}$ and find the refractive index for the left $n^{+}$ and right $n^{-}$ circular polarizations by choosing the top left and middle tensor elements respectively:
\begin{eqnarray}
n^{+}=n^{-}&=&\sqrt{\epsilon \mu - \xi_{EH}\xi_{HE} - \frac{(\xi_{EH}-\xi_{HE})^{2}\cos^{2}\theta}{4}} \nonumber \\
&+&\frac{i}{2}(\xi_{EH} - \xi_{HE})\cos \theta .
\label{refrac2}
\end{eqnarray}
Thus even in the ideal case $\epsilon \approx \mu \approx \mathbb{1}$ we do not have an isotropic index of refraction. In fact, we see that the coupling effect disappears if the fields are orthogonal ($\theta=\pi/2$) to each other.

From this analysis we see that there are two physical parameters that can be readily manipulated to achieve the refractive indices required to implement electromagnetically induced carpet cloaking.

\begin{figure}
\includegraphics[width=7cm]{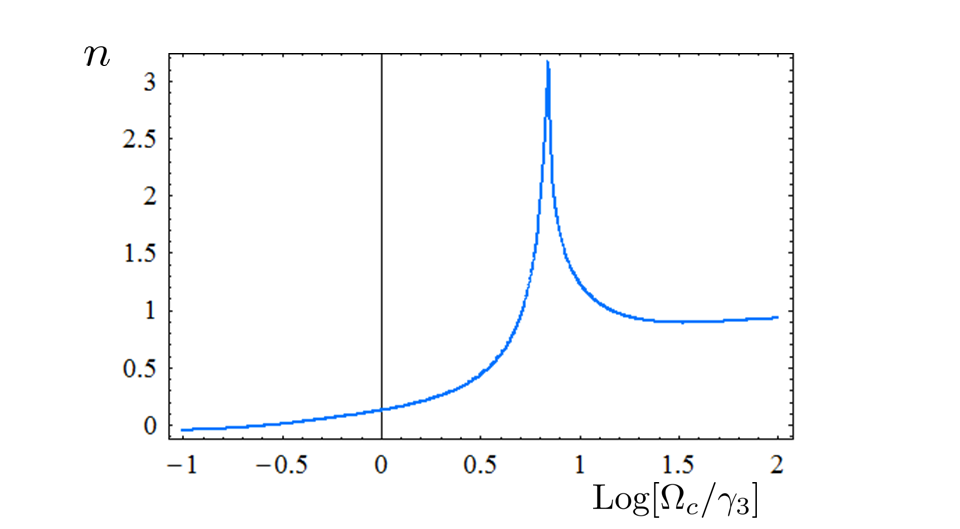}
\caption{Real part of the refractive index as a function of the coupling Rabi frequency $|\Omega_c|$ relative to $\gamma_3$, the radiative decay rate of the electric dipole transition, for $\varrho=1.56 10^{17} cm^{-3}$ and detuning $\Delta = 1.5 \gamma_p$.}
\label{refractive2}
\end{figure}

\textit{ Setting I: Refractive index $n$ as a function of cross-coupling $\Omega_c$}.
The first of our proposals to implement the carpet cloak refractive index profile relies on directly varying the coupling field $\Omega_c$ for constant detuning $\Delta = -\Delta_E = -\Delta_B = 1000$kHz and constant number of scatterers, $\varrho = 1.56^{23}$. The refractive index in a chiral medium is dependent on the relative directions of the incident laser fields \cite{Fleisch}. However, as long as we keep these directions constant, we have a simple formula, Eq.~(\ref{refrac2}), for determining $n$. Hence we can arrange our system to keep the fields at a constant relative angle of $\theta=\pi/4$ which is considerably easier to realize experimentally than if they were required to be collinear. The variation of $n$ with the coupling field is given in Figure~\ref{refractive2}.  
Using numerical analysis, we find that, at a constant detuning, a maximum refractive index $n=3.17$ for a coupling of $\Omega_c = 44$MHz can be reached. This range of $n$ is more than sufficient to implement the conformal transform given in Eq.~(\ref{zhu}). Note, however, that such a scheme requires continuous variation of the coupling field over the device which could be challenging experimentally. To avoid this, we present an alternative scheme, where $\Omega_c$ is kept constant across the cloak. 

\begin{figure}
\includegraphics[width=6cm]{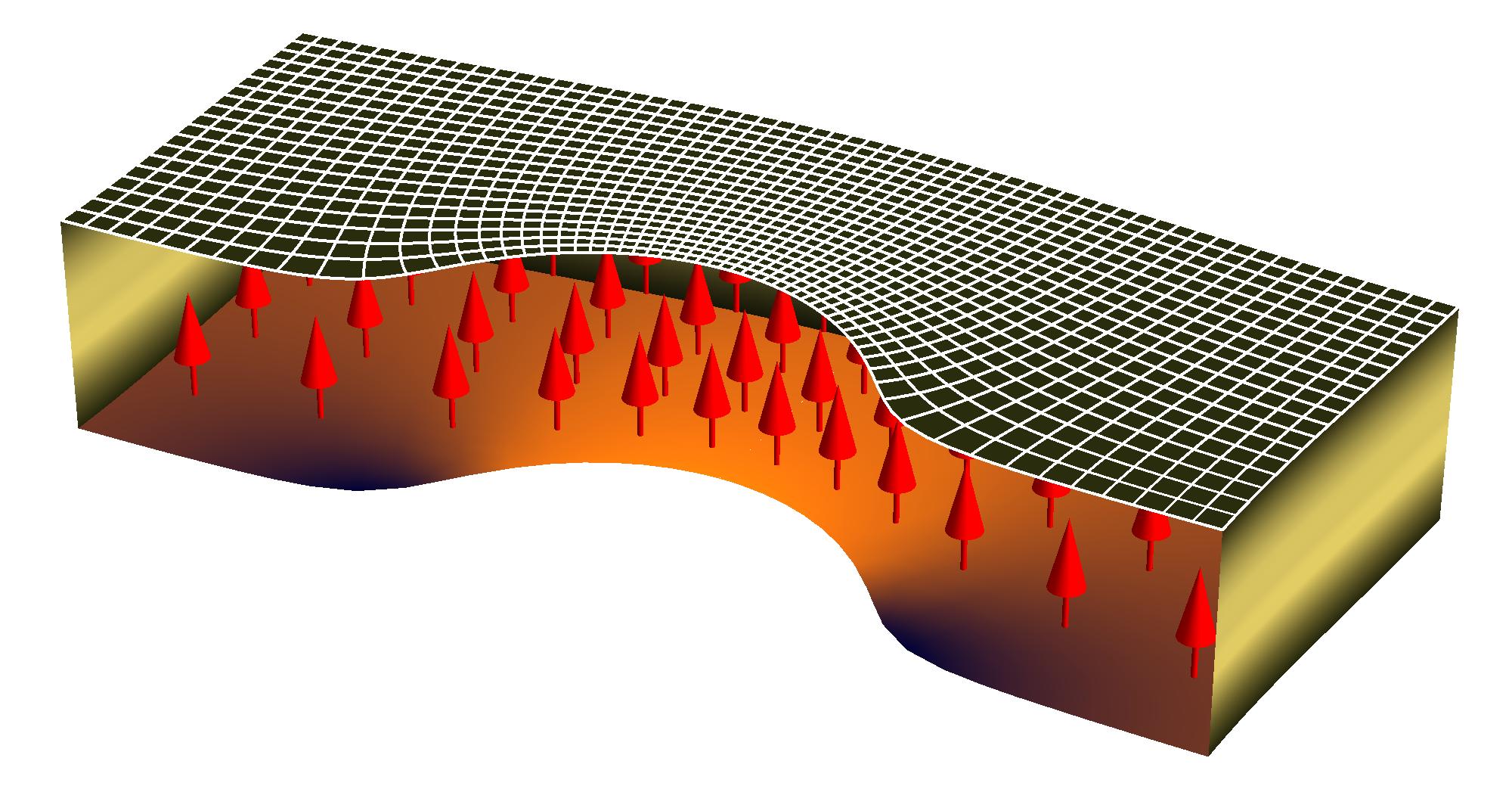}
\caption{An incident magnetic field $B$ is applied to control the detuning $\Delta$ of the chiral media implemented by a magneto-electric cross-coupling in a solid. The coupling field strength is kept at a constant value across the media. The refractive index can be tailored as in Fig.~\ref{refractive} to achieve the profiles depicted in Fig.~\ref{map}b. The insert shows the actual cloaking, the curved physical space as in Fig.~\ref{map}a, delivering desired straight lines in virtual space.}
\label{solid}
\end{figure}

\begin{figure}
\includegraphics[width=9cm]{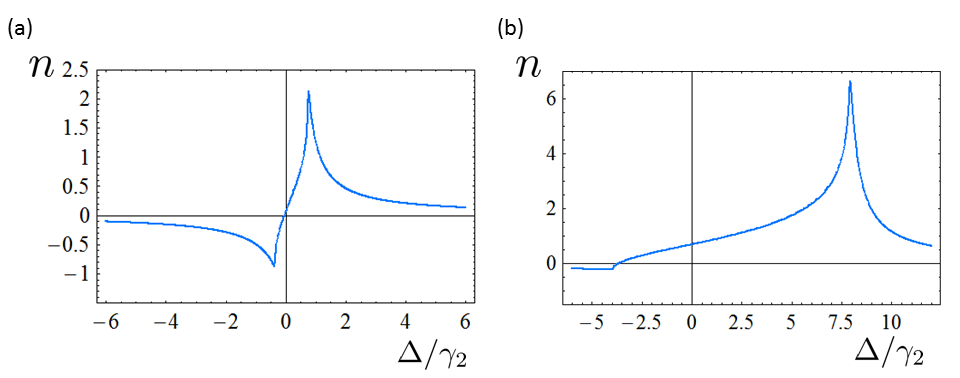}
\caption{Real parts of the refractive index including local field effects as a function of detuning $\Delta$ relative of $\gamma_2$, the radiative decay rate of the magnetic dipole transition, for two coupling field strengths: (a) $\Omega_c=16^4 \gamma_2 e^{i\pi/2}$ and (b) $\Omega_c=12^5 \gamma_2 e^{i\pi/2}$. We note that for increasing coupling strengths we can access a larger range of refractive indices but at the expense of requiring large detunings, which, at a certain magnitude become unfeasible.}
\label{refractive}
\end{figure}

\textit{ Setting II: Refractive index $n$ as a function of detuning $\Delta$}.
Here we manipulate the detuning $\Delta$ to control the refractive index of the chiral media required for a given conformal transform. The detuning is controlled by an incident gradient magnetic field that varies over the length of the device. This scheme is depicted in Figure~\ref{solid}. 
We choose a refractive index range by setting a particular value of the cross-coupling $\Omega_c$. We then vary the detuning via a magnetic field to fine tune $n$ within the selected range along the length of the cloak. The variation of refractive index with the detuning for some constant values of the cross-coupling $\Omega_c$ is plotted in Figure~\ref{refractive}. For a coupling of $\Omega_c = 15.5^4 \gamma_2 e^{i\pi/2}$ as in Fig.~\ref{refractive}(a), we find that the refractive index achieves a maximum of $n=1.83$ for a detuning $\Delta = 0.63k$Hz. Comparing with the refractive index requirements of the carpet cloak as in Fig.\ref{map}, we see that our scheme provides sufficient tunability to implement the Jukowski transform of Eq.~(\ref{zhu}). This set up is experimentally more feasible as it only relies on precise control of the magnetic field gradient (an established technique) and allows us to keep the coupling fields at a constant value. 

So far we have considered a very smooth conformal coordinate transform to implement carpet cloaking. However, such a transform limits the dimensions of the object we wish to hide. Recently, other conformal transforms have been suggested that yield larger and more irregular cloaked areas \cite{ochiai}. However, these transforms have more stringent demands on the refractive index range of the cloaking media. To test whether the chiral media are suitable to perform more demanding conformal transforms we increase the cross-coupling field (Fig.~\ref{refractive}(b)). We find that at higher field strengths we access larger refractive indices ($n=6.61$ for detuning of $\Delta = 7.935$kHz). However, these require ever larger detunings which can then become unphysical. We must also limit the coupling strength of the field to avoid thermal effects. In the ranges we have plotted, the thermal effects are extremely small as the coupling fields correspond to mW laser powers.

We have presented a scheme for electromagnetically-induced invisibility using a \textit{carpet cloak}. The scheme is based on coherent magneto-electric cross-coupling in a medium modeled by the 5-level atom interacting with a weak probe and the three control light fields (Fig.~\ref{5level}). We obtained the steady state solutions of the corresponding Liouville equation for the density matrix elements to show that such chiral media allow for a precise control of the refractive index over a wide range, while simultaneously suppressing absorption due to quantum interference effects similar to EIT. We have devised the conformal transformation that is suitable for this system and found the physical parameters required to accomplish the cloaking. 

The research leading to these results has received funding from the
European Community's Seventh Framework Programme (FP7/2007-2013) under
grant agreement $\rm n^o$ 270843 (iQIT) and has also been supported by the Scottish Universities Physics Alliance (SUPA) and by the
Engineering and Physical Sciences Research Council (EPSRC). The authors acknowledge invaluable discussions with Ulf Leonhardt and are grateful for his continuous support during this research project, as well as assistance in technical matters.


\end{document}